\documentclass{article}

\usepackage{arxiv}

\usepackage[utf8]{inputenc} 
\usepackage[T1]{fontenc}    
\usepackage{hyperref}       
\usepackage{url}            
\usepackage{booktabs}       
\usepackage{amsfonts}       
\usepackage{nicefrac}       
\usepackage{microtype}      
\usepackage{lipsum}
\usepackage{graphicx}
\usepackage{cite}
\usepackage{amsmath,amssymb,amsfonts}
\usepackage{algorithmic}
\usepackage{graphicx}
\usepackage{textcomp}
\usepackage{multirow}
\usepackage{amsmath}
\usepackage{amsfonts}
\usepackage{enumitem}
\usepackage{fancyhdr}
\usepackage{balance}
\usepackage{csquotes}
\usepackage{subcaption}
\usepackage{cleveref}
\usepackage{hyperref}
\usepackage{float}
\graphicspath{ {./images/} }

\begin{document}

\thispagestyle{empty}
{\Huge \textbf{IEEE Copyright Notice}}\\[1em]
{\large
\noindent
\textcopyright~2024 IEEE. Personal use of this material is permitted. Permission from IEEE must be obtained for all other uses, in any current or future media, including reprinting/republishing this material for advertising or promotional purposes, creating new collective works, for resale or redistribution to servers or lists, or reuse of any copyrighted component of this work in other works.
\vspace{1em}
\noindent
DOI: \href{https://doi.org/10.1109/STI64222.2024.10951092}{10.1109/STI64222.2024.10951092}
}
\newpage

\fancypagestyle{titlepage}{
  \fancyhf{}
  \fancyhead[C]{\footnotesize This work has been accepted for publication in 2024 6th International Conference on Sustainable Technologies for Industry 5.0 (STI) 14-15 December, 2024, Dhaka, Bangladesh.\\
  The final published version is available via IEEE Xplore. \\ DOI: \href{https://doi.org/10.1109/STI64222.2024.10951092}{10.1109/STI64222.2024.10951092}}
  \renewcommand{\headrulewidth}{0pt}
}

\title{Transfer Learning and Explainable AI for Brain
Tumor Classification: A Study Using MRI Data
from Bangladesh}

\author{
 Shuvashis Sarker \\
  Department of Computer Science and Engineering\\
  Ahsanullah University of Science and Technology\\
  Dhaka,Bangladesh\\
  \texttt{shuvashisofficial@gmail.com}
}

\maketitle
\thispagestyle{titlepage}

\begin{abstract}
Brain tumors, regardless of being benign or malignant, pose considerable health risks, with malignant tumors being more perilous due to their swift and uncontrolled proliferation, resulting in malignancy. Timely identification is crucial for enhancing patient outcomes, particularly in nations such as Bangladesh, where healthcare infrastructure is constrained. Manual MRI analysis is arduous and susceptible to inaccuracies, rendering it inefficient for prompt diagnosis. This research sought to tackle these problems by creating an automated brain tumor classification system utilizing MRI data obtained from many hospitals in Bangladesh. Advanced deep learning models, including VGG16, VGG19, and ResNet50, were utilized to classify glioma, meningioma, and various brain cancers. Explainable AI (XAI) methodologies, such as Grad-CAM and Grad-CAM++, were employed to improve model interpretability by emphasizing the critical areas in MRI scans that influenced the categorization. VGG16 achieved the most accuracy, attaining 99.17\%. The integration of XAI enhanced the system's transparency and stability, rendering it more appropriate for clinical application in resource-limited environments such as Bangladesh. This study highlights the capability of deep learning models, in conjunction with explainable artificial intelligence (XAI), to enhance brain tumor detection and identification in areas with restricted access to advanced medical technologies.
\end{abstract}


\section{Introduction}
Human brain is essential for controlling all bodily functions, and any disruption, such as a brain tumor, can severely affect an individual's health. Brain tumors, which can be malignant (fast-growing and invasive) or benign (slow-growing but still harmful), vary significantly in size, shape, and location. \cite{stupp2005radiotherapy} Timely identification is essential for effective therapy yet, the physical examination of MRI scans is labor-intensive and susceptible to errors owing to the extensive data and nuanced tumor margins. As a result, automation through advanced methods is necessary for efficient tumor detection.\cite{hossain2019brain}\cite{shawon2023explainable}

ML and DL methodologies have shown effective in automating brain tumor diagnosis, significantly improving diagnostic accuracy and efficiency. This work analyzed a Bangladeshi brain tumor MRI dataset, facilitating a region-specific investigation of tumor identification. This study conducts a comprehensive analysis comparing the performance of Deep neural networks (DNN) and Transfer learning (TL) models to identify which method is more effective for brain tumor classification. Furthermore, XAI methodologies, including Grad-CAM and Grad-CAM++, are utilized to show the specific regions of the MRI that the model depends on for its predictions, hence offering essential insights into model interpretability and transparency in medical contexts.
The main objectives of this study are to
\begin{enumerate}[label=\roman*]
    \item Utilizing a Bangladeshi brain tumor MRI dataset for localized research.
    \item Conducting a comparative analysis of TL models for brain tumor classification.
    \item Applying XAI techniques to visualize model focus areas for more transparent classification.
\end{enumerate}
This work seeks to improve the reliability and transparency of brain tumor identification, particularly for resource-limited settings such as Bangladesh.

\section{Related Works}
This section reviews diverse ML and DL methodologies for brain tumor diagnosis, emphasizing research conducted in Bangladesh. Brain tumor detection in Bangladesh faces challenges like limited healthcare infrastructure, high costs, unreliable internet, and a shortage of skilled professionals, along with a lack of diverse, high-quality datasets. Notwithstanding these challenges, some research have employed ML and DL to improve the precision and efficacy of detection systems, providing significant assistance in resource-constrained environments.

For instance, Hossain et al.\cite{hossain2019brain} applied Fuzzy C-Means clustering along with traditional classifiers like SVM, KNN, and CNN to extract and classify brain cancers from MRI scans, attaining a high level of accuracy of 97.87\%. Their approach demonstrated the potential of CNNs to outperform classical ML techniques, setting the stage for more advanced models. Building on this, Islam et al.\cite{islam2024improved} employed more complex DL techniques like 2D CNN and CNN-LSTM, further enhancing accuracy by utilizing ensemble methods, achieving up to 98.82\%. However, while their hybrid approach was more accurate, it introduced greater computational complexity, a challenge that would need to be addressed for practical use in resource-constrained environments.

Moving beyond traditional CNN models, Monirul et al.\cite{islam2023transfer} explored the use of TL architectures such as MobileNet, InceptionV3 and DenseNet121 achieving the highest accuracy of 99.60\%. This work demonstrated the effectiveness of TL in medical imaging, but like the earlier studies, it also highlighted the limitations of models trained on well-labeled data, as performance in real-world, noisy data could vary. To address such concerns, Shawon et al.\cite{shawon2023explainable} integrated cost-sensitive techniques with InceptionV3 and CNN models to handle imbalanced datasets, achieving 99.33\% accuracy. Their cost-sensitive approach dealt with the common issue of class imbalance in medical datasets, complementing the TL models by making the system more robust for practical use.

Focusing on the issue of computational efficiency, Majeed et al.\cite{majeed2024multi} introduced a lightweight MobileNetV3 model designed for mobile CPUs, achieving 91\% accuracy. While this model sacrificed some accuracy for efficiency, it opened the door for resource-limited settings to use these techniques in practice. Similarly, Rahman et al.\cite{rahman2024comparative} employed EfficientNetB5 for brain tumor classification, achieving 94.75\% accuracy. However, they noted the limitations of their dataset, which lacked demographic diversity, pointing to a need for more comprehensive datasets to make the model broadly applicable.

On the other hand, Khan et al.\cite{khan2020detection} approached the problem with a SVM-based method in a MATLAB environment, utilizing discrete WST for feature extraction and PCA for dimensionality reduction. Their approach, though less complex than DL, offered a simpler, more explainable method for feature-based classification, albeit with limitations in scalability for larger datasets. Building on feature-based classification, Shanjida et al.\cite{shanjida2024hybrid} proposed a lightweight CNN-SVM model that employed data augmentation and K-fold cross-validation for robust classification, achieving 96.7\% accuracy. Although their model was lightweight and efficient, the slightly lower accuracy compared to more advanced models like EfficientNet or MobileNet may limit its use in cases requiring very high precision.

Continuing the exploration of EfficientNet architectures, Manowarul et al.\cite{islam2024brainnet} employed the EfficientNet family and achieved an impressive 99.69\% accuracy with EfficientNetB3, surpassing many state-of-the-art models. Despite the high accuracy, the resource-intensive nature of EfficientNet models may limit their application in settings with limited computational power. In contrast, Sarkar et al.\cite{sarkar2023effective} used a combination of AlexNet CNN and classifiers like BayesNet, SMO, Naïve Bayes, and Random Forest, achieving a perfect accuracy of 100\% with AlexNet CNN+RF. However, the small dataset size raised concerns of potential overfitting, showing that even highly accurate models need to be tested on larger, more varied datasets to ensure generalizability.

Finally, Majib et al.\cite{majib2021vgg} proposed a stacked classifier network, VGG-SCNet, which achieved high precision, recall, and F1 scores of 99.2\%, 99.1\%, and 99.2\%, respectively. While stacking multiple classifiers increased accuracy, it also introduced computational complexity, mirroring the trade-offs seen in earlier studies that focused on ensemble models or complex DL architectures.

These studies highlight the strengths and limitations of various ML and DL approaches for brain tumor detection. High-accuracy models like CNN-LSTM, EfficientNet, and stacked classifiers often require substantial computational resources, limiting their feasibility in low-resource settings like Bangladesh. Lightweight models like MobileNet and CNN-SVM offer a balance between efficiency and accuracy but may struggle with complex cases, such as noisy or imbalanced data. Future research should focus on optimizing this balance to create models that are both accessible and accurate for resource-limited healthcare environments.

\section{MATERIALS AND METHODS}
\subsection{Dataset}
\begin{figure}[h!]
    \centering
    \includegraphics[width=\linewidth]{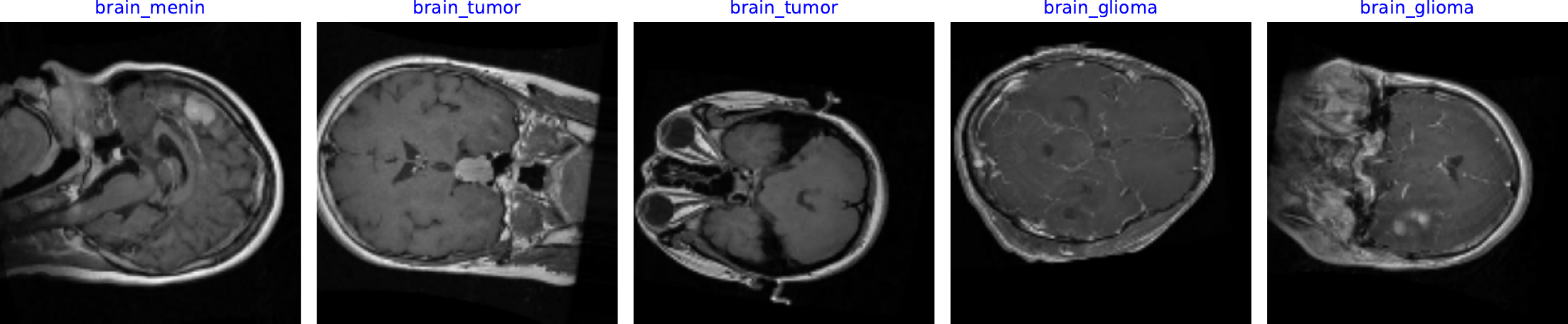}
    \caption{Sample Images of the Brain Tumour Dataset}
    \label{Figure 1}
\end{figure}

The dataset utilized for this investigation is the \textit{Bangladesh Brain Cancer MRI Dataset}, a valuable resource for advancing brain cancer research through ML and DL techniques. It consists of 6,056 MRI images categorized into three classes: \textit{\textbf{Brain Glioma}} (2,004 images), \textit{\textbf{Brain Meningioma}} (2,004 images) and \textit{\textbf{Brain Tumor}} (2,048 images), collected from various hospitals across Bangladesh to ensure a diverse and representative sample. Each image is resized to 512x512 pixels, making it compatible with a wide range of ML and DL algorithms. This dataset is essential for creating algorithms tailored to identifying and diagnosis of brain cancer within the Bangladeshi environment. The dataset is publicly available on \textit{\textbf{Mendeley}}\cite{rahman2024brain}, facilitating further research in medical diagnostics. Figure \ref{Figure 1} is showing some random samples of the dataset.
\subsection{Data Preprocessing}
The data preprocessing involves resizing the MRI images to a uniform dimension of 128x128 pixels and applying min-max normalization. This normalization adjusts pixel values based on the minimum and maximum intensity values in each image, scaling them between 0 and 1. This ensures consistency across the dataset, making it well-prepared for further analysis or model training.
\subsection{Dataset Splitting}
The dataset is partitioned into training, validation and test sets in an 80:10:10 ratio. 80\% of the data is designated for model training, 10\% for validation to optimize hyperparameters, and the final 10\% is assigned for testing to assess the model's performance on novel data. This division guarantees a balanced training methodology, mitigating overfitting and enhancing the model's adaptation abilities.

\subsection{Data Augmentation}
Data augmentation is crucial for enhancing dataset variety and boosting model generalization by mitigating overfitting, particularly when the dataset is constrained. The augmentation process involves several transformations to enhance variation in the images.First, horizontal flipping is applied, which mirrors the image along the vertical axis. Gaussian noise is then added, generated using a normal distribution with a standard deviation of 0.0023, and the pixel values are clipped between 0 and 1 to maintain valid intensity ranges. Next, contrast is adjusted by scaling the pixel values by 0.79, followed by increasing the brightness by adding 0.24 to the contrast-adjusted image. Finally, random rotation is applied, where the image is rotated by a randomly selected angle from the set \{-13°, -9°, +9°, +13°\}. These augmentations help the model learn more robust features by exposing it to diverse transformations of the same image, thus enhancing its adaptability to unseen data.
\begin{figure}[ht]
    \centering
   \includegraphics[width=0.6\linewidth]{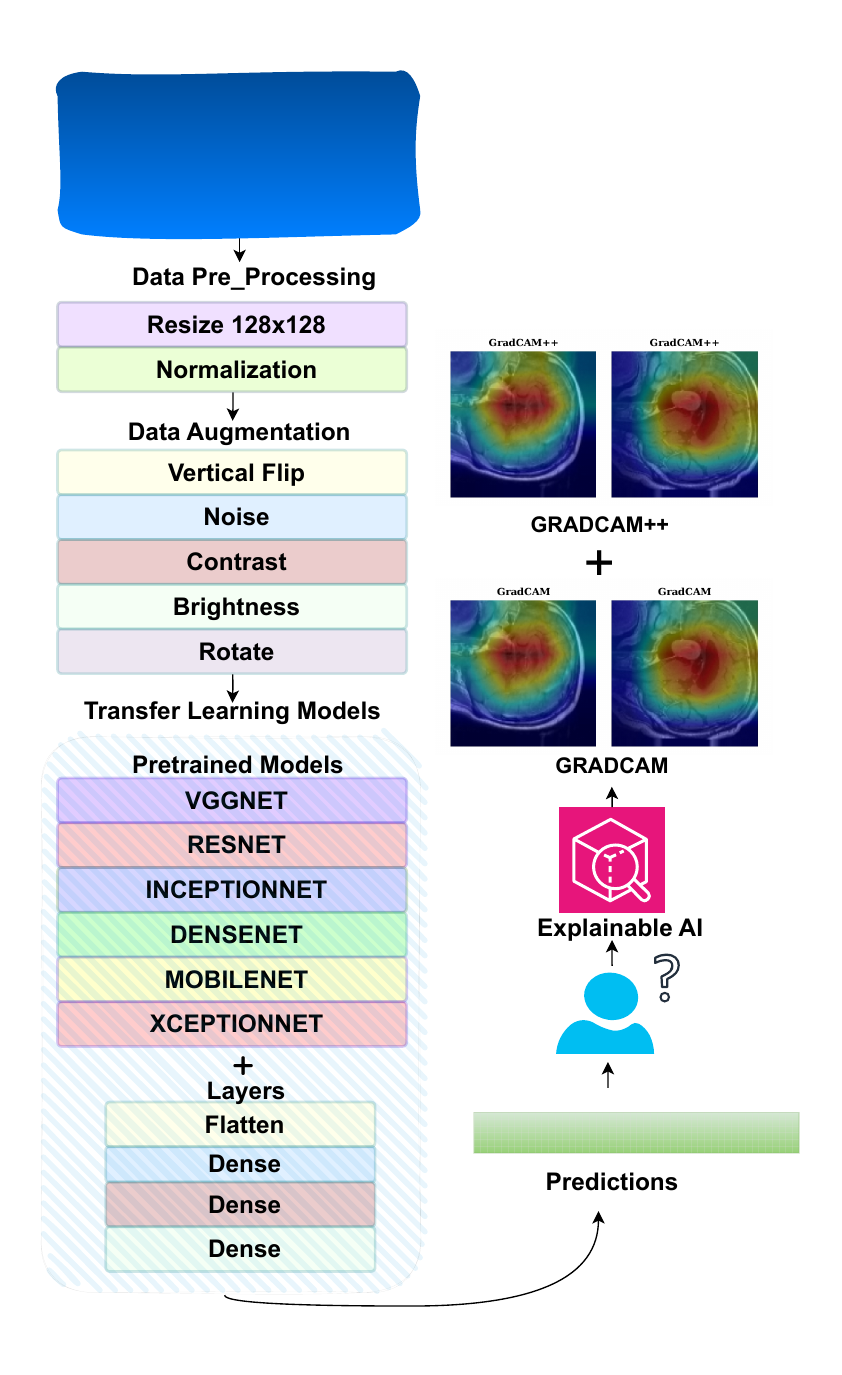}
    \caption{Proposed Model Architecture}
    \label{Figure 2}
\end{figure}

\subsection{Proposed Model Architecture}
The proposed model of this paper utilizes \textit{TL} by incorporating multiple \textit{pretrained models}, including \textit{VGGNet, ResNet, InceptionNet, DenseNet, MobileNet} and \textit{XceptionNet}. Each of these models is initialized with \textit{ImageNet} weights and fine-tuned for the specific classification task. The pretrained models act as feature extractors by using their convolutional layers, and additional custom layers are added to handle the classification. By adding new layers on top of these pretrained models and fine-tuning their weights, they effectively become \textit{TL models}, enabling them to apply learned knowledge to the new dataset. After extracting features from the pretrained models, the architecture adds layers for \textit{flattening} the feature map, followed by dense layers with \textit{dropout} to prevent overfitting. Finally, the output layer is designed for multi-class classification, predicting the target categories. Figure \ref{Figure 2} shows the full model architecture, illustrating how TL and custom layers are combined. This approach ensures that the model benefits from rich, pre-learned feature representations while adapting to the specific problem through custom layers and adjustments. The model undergoes training for a designated number of epochs, with the objective of enhancing accuracy using suitable loss and optimization functions.

\section{Result Analysis}
The processes were executed on a machine equipped with an AMD Ryzen 5 3600 processor, an NVIDIA GeForce RTX 3060Ti 8GB GPU, and 32GB of DDR4 RAM. The model implementation was based on Keras \cite{chollet2015keras}, an API running on TensorFlow \cite{tensorflow2015-whitepaper}. Throughout the experiments, the model was trained for an average of 30 epochs, and the learning rate was fine-tuned between 1e-3 and 1e-5, with the optimal results selected for further analysis. The cost function used was sparse categorical cross-entropy, and the Adam optimizer was employed, with its performance compared against other optimizers, such as Adagrad and SGD. Adam demonstrated superior performance in achieving the best results, validating its effectiveness in this particular task.
\begin{table}[ht]
\centering
\renewcommand{\arraystretch}{1.5}
\caption{Performance comparison of different models}
\label{tab:performance_comparison}
\begin{tabular}{|c|c|c|c|c|c|}
\hline
\textbf{Model}                     & \textbf{Disease} & \textbf{Accuracy}              & \textbf{Precision} & \textbf{Recall} & \textbf{F1 Score} \\ \hline
\multirow{3}{*}{\textbf{VGG16}}    & Glioma           & \multirow{3}{*}{\textbf{0.99}} & 0.98               & 0.98            & 0.98              \\ \cline{2-2} \cline{4-6} 
                                   & Menin            &                                & 0.98               & 0.98            & 0.98              \\ \cline{2-2} \cline{4-6} 
                                   & Tumor            &                                & 0.97               & 0.98            & 0.98              \\ \hline
\multirow{3}{*}{\textbf{VGG19}}    & Glioma           & \multirow{3}{*}{0.98}          & 1.00               & 0.99            & 0.99              \\ \cline{2-2} \cline{4-6} 
                                   & Menin            &                                & 0.98               & 0.97            & 0.98              \\ \cline{2-2} \cline{4-6} 
                                   & Tumor            &                                & 0.97               & 0.99            & 0.98              \\ \hline
\multirow{3}{*}{\textbf{DenseNet}} & Glioma           & \multirow{3}{*}{0.98}          & 0.99               & 0.98            & 0.98              \\ \cline{2-2} \cline{4-6} 
                                   & Menin            &                                & 0.98               & 0.98            & 0.98              \\ \cline{2-2} \cline{4-6} 
                                   & Tumor            &                                & 0.97               & 0.98            & 0.98              \\ \hline
\multirow{3}{*}{\textbf{InceptionResNetV2}} & Glioma           & \multirow{3}{*}{0.95}          & 0.97               & 0.98            & 0.97              \\ \cline{2-2} \cline{4-6} 
                                   & Menin            &                                & 0.94               & 0.94            & 0.94              \\ \cline{2-2} \cline{4-6} 
                                   & Tumor            &                                & 0.95               & 0.94            & 0.95              \\ \hline
\multirow{3}{*}{\textbf{MobileNetV2}} & Glioma           & \multirow{3}{*}{0.96}          & 0.96               & 0.97            & 0.96              \\ \cline{2-2} \cline{4-6} 
                                   & Menin            &                                & 0.96               & 0.97            & 0.96              \\ \cline{2-2} \cline{4-6} 
                                   & Tumor            &                                & 0.96               & 0.99            & 0.98              \\ \hline
\multirow{3}{*}{\textbf{Xception}} & Glioma           & \multirow{3}{*}{0.97}          & 0.99               & 0.98            & 0.98              \\ \cline{2-2} \cline{4-6} 
                                   & Menin            &                                & 0.97               & 0.94            & 0.95              \\ \cline{2-2} \cline{4-6} 
                                   & Tumor            &                                & 0.94               & 0.99            & 0.96              \\ \hline
\multirow{3}{*}{\textbf{ResNet50V2}} & Glioma           & \multirow{3}{*}{0.95}          & 0.97               & 0.95            & 0.96              \\ \cline{2-2} \cline{4-6} 
                                   & Menin            &                                & 0.94               & 0.94            & 0.94              \\ \cline{2-2} \cline{4-6} 
                                   & Tumor            &                                & 0.95               & 0.97            & 0.96              \\ \hline
\multirow{3}{*}{\textbf{InceptionV3}} & Glioma           & \multirow{3}{*}{0.96}          & 0.97               & 0.98            & 0.97              \\ \cline{2-2} \cline{4-6} 
                                   & Menin            &                                & 0.96               & 0.94            & 0.95              \\ \cline{2-2} \cline{4-6} 
                                   & Tumor            &                                & 0.95               & 0.97            & 0.96              \\ \hline
\end{tabular}%
\end{table}

\par The Table \ref{tab:performance_comparison} presents a summary of the efficacy of different DL models in the classification of three brain illness types: Glioma, Meningioma, and Tumor. Among the models, VGG16 stands out with the highest accuracy of 99.17\%, making it the most effective for this task. Other models, such as VGG19 and DenseNet, also show strong performance, achieving accuracy levels of 98\%. Although some models have slightly lower accuracies, ranging from 95\% to 96\%, they still perform competitively. 
\begin{figure}[ht]
    \centering
    \begin{subfigure}{0.8\linewidth}
        \centering
        \includegraphics[width=\linewidth]{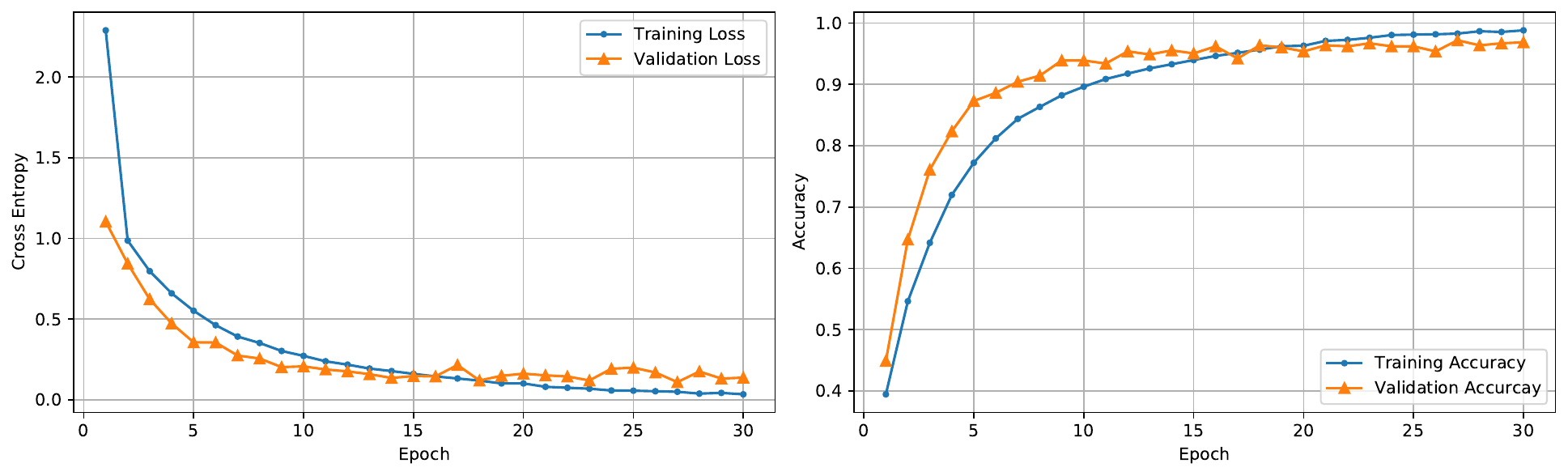}
        \caption{VGG16 Training Curves}
    \end{subfigure}%
    \hspace{0.04\textwidth} 
    \begin{subfigure}{0.8\linewidth}
        \centering
        \includegraphics[width=\linewidth]{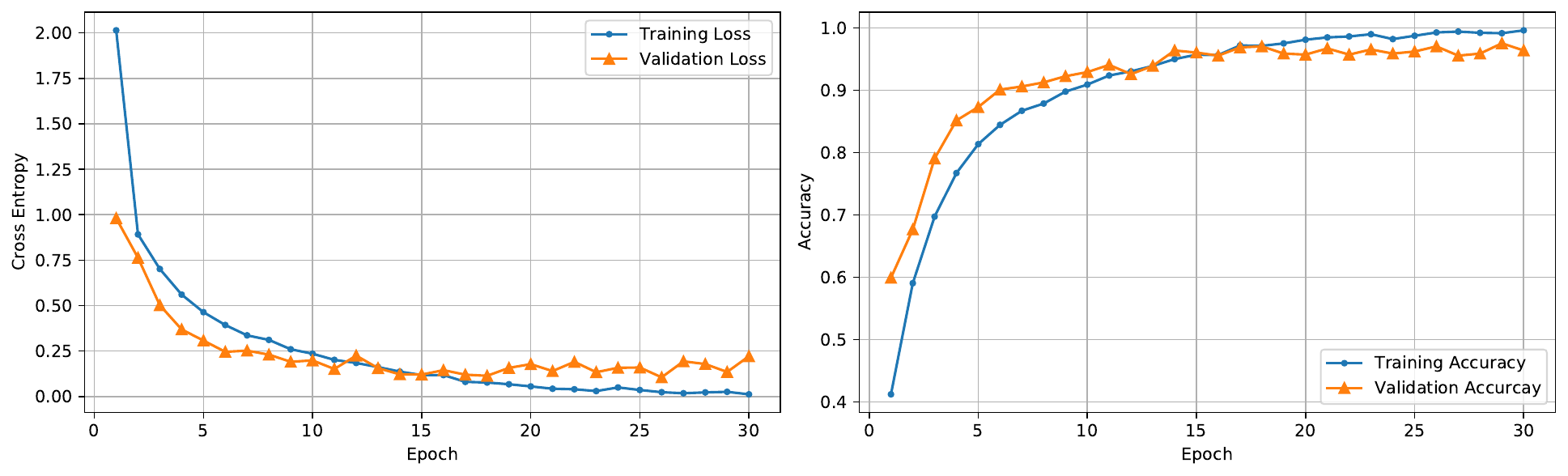}
        \caption{VGG19 Training Curves}
    \end{subfigure}
    \caption{Training and validation curves.}
    \label{fig:training_curves}
\end{figure}
\begin{figure}[H]
    \centering
    \begin{subfigure}{0.8\linewidth} 
        \centering
        \includegraphics[width=\linewidth]{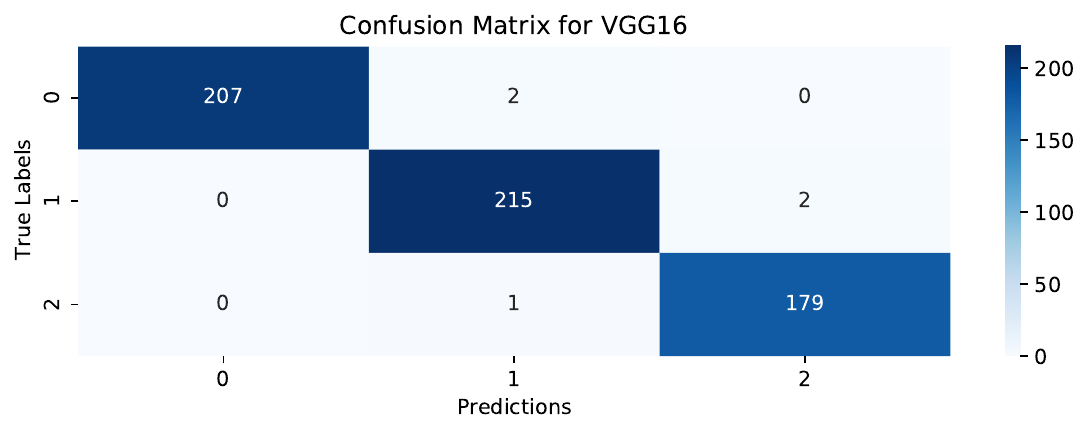}
        \caption{VGG16 Confusion Matrix}
    \end{subfigure}%
    \hspace{0.05\textwidth} 
    \begin{subfigure}{\linewidth} 
        \centering
        \includegraphics[width=0.8\linewidth]{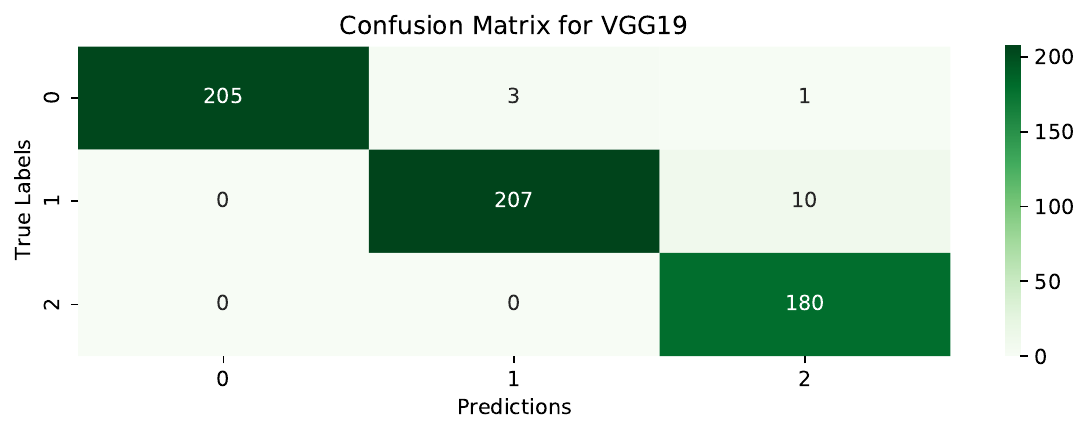}
        \caption{VGG19 Confusion Matrix}
    \end{subfigure}
    \caption{Confusion matrices of different Models.}
    \label{fig:confusion_matrices}
\end{figure}
The training curves for VGG16 and VGG19 exhibit a steady decline in training and validation loss, accompanied by a rise in training and validation accuracy across the epochs. This indicates that both models efficiently reduce error and enhance forecast accuracy over time. Figure \ref{fig:training_curves} illustrates the trends in both accuracy and loss metrics throughout the training process, which was conducted for an average of 30 epochs. This early stopping approach helps prevent overfitting, as extending the training to a larger number of epochs could potentially lead to the model learning noise in the training data, resulting in decreased generalizability. By stopping at 30 epochs, the model achieves a balance between learning the underlying patterns and avoiding overfitting, ensuring robust performance on unseen data. Following the training analysis, Figure \ref{fig:confusion_matrices} presents the confusion matrices for the VGG16 and VGG19 models, which illustrate their performance in classifying the three classes of brain diseases.In both matrices, most predictions are accurately classified, as evidenced by the elevated values along the diagonal. The VGG16 model shows slightly better performance in some classes, particularly with fewer misclassifications in Class 1 compared to VGG19. Overall, VGG16 emerges as the top-performing model in terms of accuracy, highlighting its suitability for brain disease classification. 


\begin{table}[ht]
\renewcommand{\arraystretch}{2} 
\centering
\caption{Comparison of Related Work in Brain Tumor Detection Approaches}
\begin{tabular}{|p{3cm}|p{5cm}|p{4cm}|c|p{0.8cm}|} 
\hline
\textbf{Authors}              & \textbf{Dataset}                                          & \textbf{Approach}                                                      & \textbf{Accuracy} & \textbf{Use of XAI} \\ \hline
Hossain et al. \cite{hossain2019brain}  & BRATS Dataset                        & \raggedright Fuzzy C-Means + SVM, KNN, CNN                                  & 97.87\%           & No                  \\ \hline
Islam et al. \cite{islam2024improved}   & \raggedright Br35H, Brain Tumor Detection, SARTAJ Brain Tumor Classification (MRI), and Radiya Brain Tumor Dataset & \raggedright 2D CNN, CNN-LSTM + ensemble methods                            & 98.82\%           & No                  \\ \hline
Monirul et al. \cite{islam2023transfer} & Kaggle Brain Tumor MRI Dataset (MasoudNickparvar)  & \raggedright TL (MobileNet, InceptionV3, DenseNet121)        & 99.60\%           & No                  \\ \hline
Shawon et al. \cite{shawon2023explainable} & \raggedright Br35H, Brain Tumor Detection and Brain MRI Images for Tumor Detection & \raggedright Cost-sensitive learning with InceptionV3, CNN                 & 99.33\%           & Yes                 \\ \hline
Majeed et al. \cite{majeed2024multi}    & Brain Tumor MRI Images (44 Classes, Private Collection) & \raggedright Lightweight MobileNetV3 for mobile CPUs                        & 91.00\%           & No                  \\ \hline
Rahman et al. \cite{rahman2024comparative} & MRI Brain Tumor Dataset (44 Classes) & \raggedright EfficientNetB5 for brain tumor classification                & 94.75\%           & No                  \\ \hline
Khan et al. \cite{khan2020detection}    & Cancer Imaging Archive                & \raggedright SVM with discrete wavelet transform + PCA                      & 94\%               & No                  \\ \hline
Shanjida et al. \cite{shanjida2024hybrid} & Figshare Brain Tumor Dataset          & \raggedright Lightweight CNN-SVM + K-fold cross-validation                & 96.70\%           & No                  \\ \hline
Manowarul et al. \cite{islam2024brainnet} & Publicly Available Contrast-Enhanced MRI Dataset & \raggedright EfficientNetB3 for brain tumor classification                 & 99.69\%           & No                  \\ \hline
Majib et al. \cite{majib2021vgg}        & \raggedright Kaggle Dataset (Training) and Pathology Institute in Bangladesh (Test) & \raggedright Stacked classifier network (VGG-SCNet)                         & 99.20\%           & No                  \\ \hline
\textbf{Proposed Approach}        & \raggedright Bangladesh Brain Cancer MRI Dataset & \raggedright \textbf{Advanced CNN models with Grad-CAM/Grad-CAM++}                 & \textbf{99.17\%}  & \textbf{Yes}        \\ \hline
\end{tabular}
\label{tab:related_work_comparison}
\end{table}

Overall, VGG16 demonstrates exceptional accuracy in brain tumor classification, making it a dependable model for clinical use. This study highlights the effectiveness of DL models in addressing the challenges of brain tumor detection in Bangladesh, where timely diagnosis is essential to improve patient outcomes. Table \ref{tab:related_work_comparison} compares various approaches, from classical methods like Fuzzy C-Means with SVM and KNN to advanced DL models such as CNN-LSTM and transfer learning. Although some approaches show slightly higher accuracy, the novelty of this work lies in its use of localized MRI data from Bangladesh and multiple explainable AI (XAI) techniques, specifically Grad-CAM and Grad-CAM++. Unlike studies that rely on general datasets, this research addresses the unique healthcare needs in Bangladesh, enhancing the model's adaptability for this population. With an accuracy of 99.17\% and dual-XAI interpretability, this approach not only ensures high precision but also transparency, enabling clinicians to understand and validate diagnoses. This focus on localized data and dual-XAI integration distinguishes this study as a practical, interpretable AI tool suited for resource-limited settings where trust in AI is critical.

\section{Explainable AI(XAI)}
XAI includes techniques designed to enhance the transparency and interpretability of AI model decisions, hence making complex models more understandable to humans. Techniques like Grad-CAM and Grad-CAM++ highlight image areas that significantly influence a model's output, helping users see which data aspects affect its predictions.

Grad-CAM and Grad-CAM++ are interpretability methodologies extensively employed in deep learning to visualize the critical areas of an input image that substantially influence a model's classification result. Grad-CAM \cite{selvaraju2020grad} functions by calculating the gradient of the class score \( Y^c \) with respect to the feature maps \( A^k \). The gradients are globally averaged to obtain significant weights \( \alpha_k^c \) for each feature map \( A^k \). The resultant weighted sum of the feature maps undergoes a ReLU activation to generate a heatmap that emphasizes the significant areas of the image. The equation for Grad-CAM \cite{faria2024explainable} can be written as:

\[
\alpha_k^c = \frac{1}{Z} \sum_i \sum_j \frac{\partial Y^c}{\partial A^k_{ij}}
\]
\[
L_{\text{Grad-CAM}}^c = \text{ReLU}\left( \sum_k \alpha_k^c A^k \right)
\]

Where:
\begin{itemize}
    \item \( \alpha_k^c \) is the importance weight for the \( k \)-th feature map of class \( c \).
    \item \( Z \) is the total number of pixels in the feature map.
    \item \( \frac{\partial Y^c}{\partial A^k_{ij}} \) is the gradient of the class score \( Y^c \) with respect to activation \( A^k \) at coordinates \( (i, j) \).
    \item The ReLU function is used to emphasize positive contributions only.
\end{itemize}

Grad-CAM++ \cite{chattopadhay2018grad} enhances this process by incorporating second-order derivatives, allowing for more precise localization of important regions in the image. Grad-CAM++ refines the weighting scheme, taking into account both first-order and second-order gradients for greater interpretability, especially when multiple objects or patterns are present in the image. The formula for Grad-CAM++ is expressed as:

\[
\alpha_k^c = \frac{1}{Z} \sum_i \sum_j \left( \frac{\partial^2 Y^c}{\partial (A_{ij}^k)^2} + 2 \cdot \frac{\partial Y^c}{\partial A_{ij}^k} \right)
\]
\[
L_{\text{Grad-CAM++}}^c = \text{ReLU}\left( \sum_k \alpha_k^c A^k \right)
\]

\begin{figure}[ht]
    \centering

    \begin{subfigure}{0.24\textwidth}
        \centering
        \includegraphics[width=\textwidth]{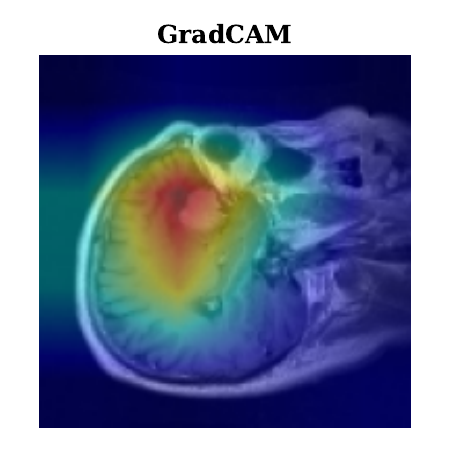}
        \caption{Menin}
    \end{subfigure}
    \begin{subfigure}{0.24\textwidth}
        \centering
        \includegraphics[width=\textwidth]{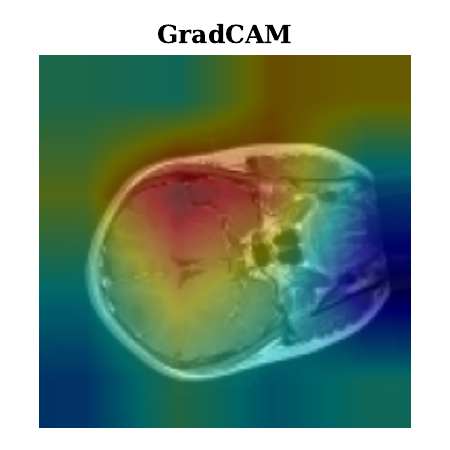}
        \caption{Glioma}
    \end{subfigure}
    \begin{subfigure}{0.24\textwidth}
        \centering
        \includegraphics[width=\textwidth]{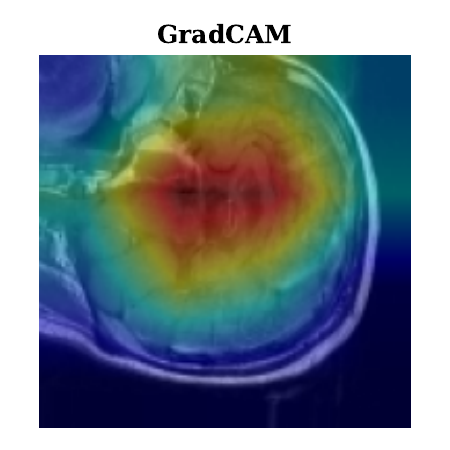}
        \caption{Tumor}
    \end{subfigure}
    \begin{subfigure}{0.24\textwidth}
        \centering
        \includegraphics[width=\textwidth]{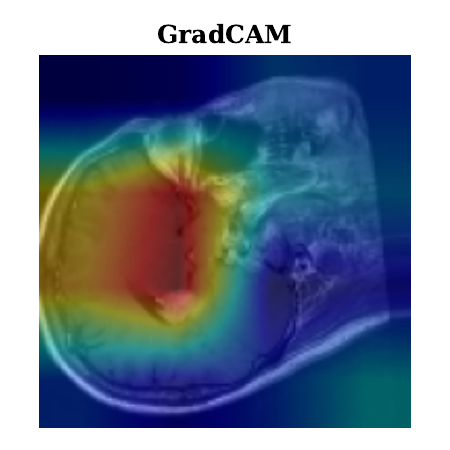}
        \caption{Menin}
    \end{subfigure}

    \vspace{0.3cm}

    \begin{subfigure}{0.24\textwidth}
        \centering
        \includegraphics[width=\textwidth]{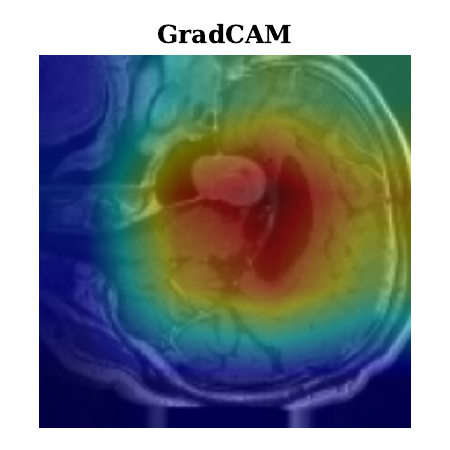}
        \caption{Tumor}
    \end{subfigure}
    \begin{subfigure}{0.24\textwidth}
        \centering
        \includegraphics[width=\textwidth]{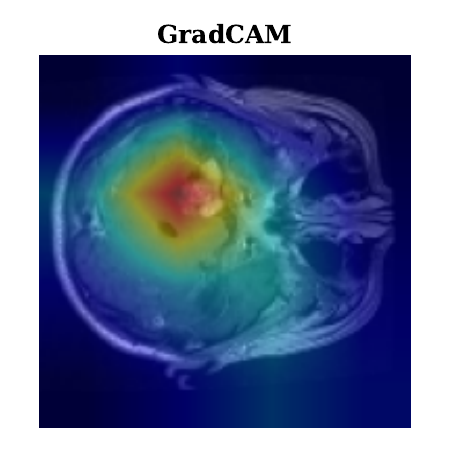}
        \caption{Menin}
    \end{subfigure}
    \begin{subfigure}{0.24\textwidth}
        \centering
        \includegraphics[width=\textwidth]{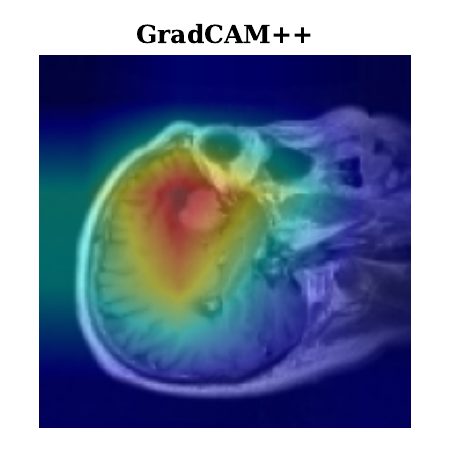}
        \caption{Menin (++)}
    \end{subfigure}
    \begin{subfigure}{0.24\textwidth}
        \centering
        \includegraphics[width=\textwidth]{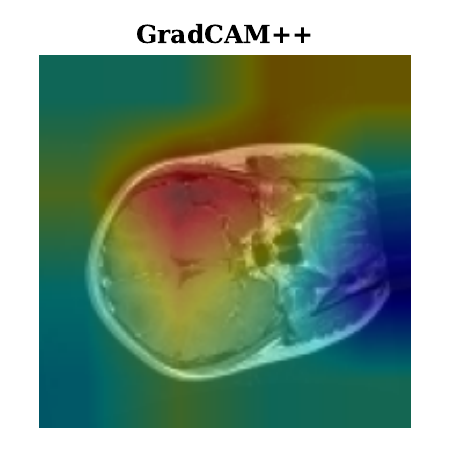}
        \caption{Glioma (++)}
    \end{subfigure}

    \vspace{0.3cm}

    \begin{subfigure}{0.24\textwidth}
        \centering
        \includegraphics[width=\textwidth]{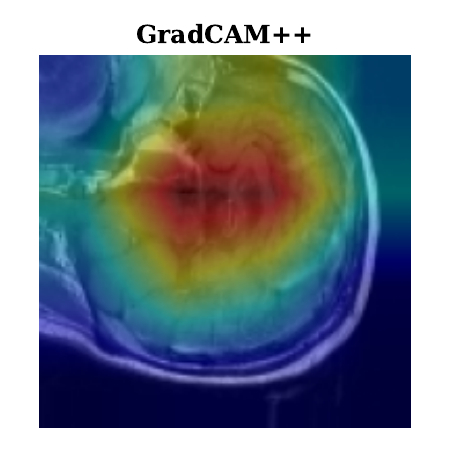}
        \caption{Tumor (++)}
    \end{subfigure}
    \begin{subfigure}{0.24\textwidth}
        \centering
        \includegraphics[width=\textwidth]{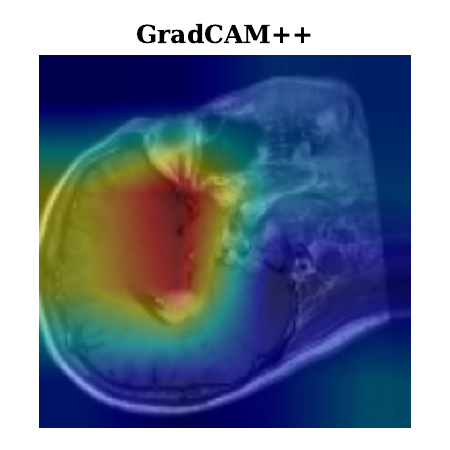}
        \caption{Menin (++)}
    \end{subfigure}
    \begin{subfigure}{0.24\textwidth}
        \centering
        \includegraphics[width=\textwidth]{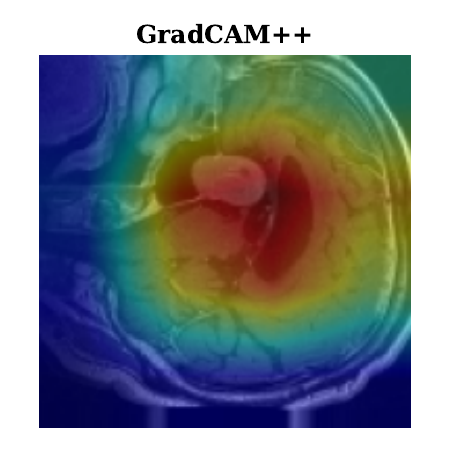}
        \caption{Tumor (++)}
    \end{subfigure}
    \begin{subfigure}{0.24\textwidth}
        \centering
        \includegraphics[width=\textwidth]{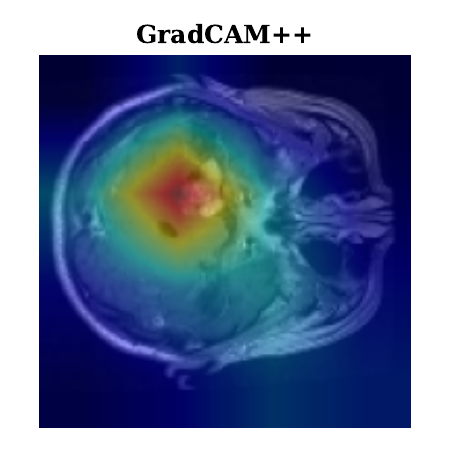}
        \caption{Menin (++)}
    \end{subfigure}

    \caption{Grad-CAM and Grad-CAM++ visualizations of brain tumour region identification across different models and tumour types.}
    \label{fig:gradcam_visualizations}
\end{figure}

By leveraging these higher-order gradients, Grad-CAM++ generates a more detailed and accurate heatmap, making it particularly effective for images with multiple objects or complex patterns, offering refined interpretability compared to Grad-CAM. Figure \ref{fig:gradcam_visualizations} demonstrates the utilization of Grad-CAM and Grad-CAM++ in our research, augmenting the interpretability of brain tumor classifications by offering critical insights into the decision-making processes of our deep learning models for tumor types including Menin, Glioma, and Tumor.  Both techniques consistently capture similar critical regions, focusing on areas around the tumor’s core where distinguishing features are most prominent, thus supporting the reliability of model predictions across different tumor types.

 Grad-CAM++ further refines these regions with more precise localization, particularly in complex cases like Glioma, where subtle boundaries exist between tumor and healthy tissue. This overlap in identified regions validates their importance for classification, with Grad-CAM++ offering a finer resolution that reveals deeper insights into tumor morphology. Comparing these XAI techniques provides a clearer understanding of feature importance by highlighting both similar and distinct regions, which strengthens interpretability and model transparency. This approach not only validates the model's predictions but also enhances trust in AI as a precise and reliable tool in complex medical diagnostics, reinforcing its value in understanding tumor structures.

\section{Conclusion}
\balance
In conclusion, This study highlights the unique contribution of combining deep learning with explainable AI techniques, specifically using VGG16 enhanced by Grad-CAM and Grad-CAM++, for brain tumor classification within the context of Bangladesh’s healthcare constraints. While advanced deep learning models can achieve high accuracy, the addition of XAI approaches are crucial for this application, as they furnish doctors with transparent insights into the model's decision-making process. This visual interpretability is crucial for AI deployment in clinical settings, where trust and transparency are paramount. The study goes beyond pure accuracy by emphasizing transparency, addressing practical challenges such as limited computational resources and the lack of specialized datasets tailored to Bangladesh’s needs. By pioneering the use of localized MRI data from Bangladesh and incorporating interpretable models, this work is particularly suited for real-world application in resource-constrained environments. Future work will build on this foundation by gathering more diverse local data and exploring transformer-based models to further enhance the model’s robustness and applicability, positioning this study as a pivotal step toward scalable, interpretable AI solutions in healthcare.

\bibliographystyle{unsrt} 
\bibliography{references.bib}

\end{document}